# A new strain energy function for modelling ligaments and tendons whose fascicles have a helical arrangement of fibrils

Tom Shearer *

School of Mathematics, University of Manchester, Manchester M13 9PL, United Kingdom

ABSTRACT

A new strain energy function for the hyperelastic modelling of ligaments and tendons whose fascicles have a helical arrangement of fibrils is derived. The stress–strain response of a single fascicle whose fibrils exhibit varying levels of crimp throughout its radius is calculated and used to determine the form of the strain energy function. The new constitutive law is used to model uniaxial extension test data for human patellar tendon and is shown to provide an excellent fit, with the average relative error being 9.8%. It is then used to model shear and predicts that the stresses required to shear a tendon are much smaller than those required to uniaxially stretch it to the same strain level. Finally, the strain energy function is used to model ligaments and tendons whose *fascicles* are helical, and the relative effects of the fibril helix angle, the fascicle helix angle and the fibril crimp variable are compared. It is shown that they all have a significant effect; the fibril crimp variable governs the non-linearity of the stress–strain curve, whereas the helix angles primarily affect its stiffness. Smaller values of the helix angles lead to stiffer tendons; therefore, the model predicts that one would expect to see fewer helical sub-structures in stiff positional tendons, and more in those that are required to be more flexible.

## 1. Introduction

Ligaments and tendons are important connective tissues; ligaments connect bone to bone, providing stability and allowing joints to function correctly, and tendons connect muscle to bone to transfer forces generated by muscles to the skeleton. They both have a hierarchical structure consisting of several fibrous subunits (Kastelic et al., 1978; Screen et al., 2004), which, from largest to smallest, can be defined as follows: fascicles (50–400 μm diameter), fibrils (50–500 nm), sub-fibrils (10–20 nm), microfibrils (3–5 nm), and finally, the tropocollagen molecule (∼1.5 nm). The geometrical arrangement of many of these subunits varies between different ligaments and tendons; for example, the patellar tendon's fascicles are coaligned with its longitudinal axis, whereas the anterior cruciate ligament's are helical (Shearer et al., 2014). The fibrils within a fascicle may also either be coaligned or helical with respect to its longitudinal axis (Yahia and Drouin, 1989). In both cases, the fibrils exhibit an additional waviness, called crimp, which is superimposed upon their average direction and varies in magnitude throughout the fascicle's radius (Kastelic et al., 1978; Yahia and Drouin, 1989). This intricate structure produces complex mechanical behaviour such as anisotropy, viscoelasticity and non-linearity, which varies between different ligaments and tendons (Benedict et al., 1968; Tipton et al., 1986). It is not currently known, however, which levels of the hierarchy are most influential in governing their mechanical performance.

To begin understanding these mechanical features, it is of interest to model their *elastic* properties, neglecting viscoelasticity. Elastic models are expected to be valid in both the low and extremely high strain rate limits where hysteresis is minimised. Ligament and tendon stress–strain behaviour under uniaxial tension is characterised by an initial non-linear region of increasing stiffness, termed the *toe-region*, followed by a linear region before the onset of failure (Fig. 1). Several authors have derived expressions to describe this behaviour (Frisen et al., 1969; Kastelic et al., 1980; Kwan and Woo, 1989); however, to consider more complex deformations, it is useful to characterise the elasticity of a material in terms of a strain energy function (SEF).

Many non-linear elastic SEFs have been proposed for soft tissues (Fung, 1967; Gou, 1970; Holzapfel et al., 2000), but few have focused specifically on ligaments and tendons. Whilst many SEFs are general enough to be applied to modelling ligaments and tendons, the majority of them contain variables that cannot be directly experimentally measured (Shearer, 2015). This limits their ability to analyse which physical quantities are most important in governing a specific ligament's or tendon's behaviour. Microstuctural models are better equipped to facilitate this analysis, provided their parameters can be experimentally determined.

* Tel.: +44 161 275 5810.
E-mail address: tom.shearer@manchester.ac.uk

**Nomenclature**

| | |
|---|---|
| $\theta_p(\rho)$ | fibril crimp angle distribution |
| $p$ | crimp angle distribution parameter |
| $\rho$ | non-dimensional radial variable in fascicle |
| $\theta_o$ | crimp angle of outermost fibrils |
| **P**, **p** | unit vectors in fibril direction before/after fascicle stretch |
| $\alpha, \psi$ | fibril/fascicle helix angle |
| $\lambda, \epsilon$ | given fascicle stretch/strain |
| $\Lambda$ | component of $\lambda$ in fibril direction |
| $\Lambda_p(\rho)$ | stretch in fibril direction as fibrils at radius $\rho$ become taut |
| $\Lambda^*, \lambda^*$ | stretch in fibril/fascicle direction that tautens outer fibrils |
| $\epsilon^*$ | critical fascicle strain as outer fibrils become taut |
| $R_p$ | radius within which all fibrils are taut for a given $\lambda$ |
| $P_p$ | tensile load experienced by fascicle |
| $\sigma_p(\rho)$ | contribution of fibril stress at radius $\rho$ in fascicle direction |
| $\sigma_p^f(\rho), \epsilon_p^f(\rho)$ | stress/strain in fibril at radius $\rho$ |
| $E$ | fibril Young's modulus |
| $\tau_p$ | average traction in fascicle direction |
| $\beta$ | $2(1 - \cos^3 \theta_o)/(3 \sin^2 \theta_o)$ |
| $W$ | strain energy function |
| $I_1, I_4$ | isotropic/anisotropic strain invariant |
| $\phi$ | collagen volume fraction |
| $W_m, W_f$ | component of $W$ associated with matrix/fibrils |
| **B**, **C** | left/right Cauchy–Green tensor |
| **M**, **m** | undeformed/deformed fascicle direction vector |
| **F** | deformation gradient |
| **T** | Cauchy stress |
| $Q$ | Lagrange multiplier |
| $\mathbf{T}_f, \mathbf{t}_f$ | component of stress/traction associated with fascicles |
| $\hat{\mathbf{m}}$ | unit vector in direction of **m** |
| $\gamma, \eta$ | constants defined in Eqs. (32) and (33) |
| $\mu$ | ground state shear modulus of ligament/tendon matrix |
| $\hat{\eta}$ | constant defined in Eq. (41) |
| $R, \Theta, Z$ | circular cylindrical coordinates in undeformed configuration |
| $A, a$ | undeformed/deformed tendon radius |
| $L, l$ | undeformed/deformed tendon length |
| $r, \theta, z$ | circular cylindrical coordinates in deformed configuration |
| $\zeta$ | stretch in longitudinal direction of tendon |
| $\mathbf{e}_i, \mathbf{E}_j$ | basis vectors in deformed/undeformed configuration |
| **n** | outer unit normal to curved surface of tendon |
| $S_{zz}, S_{zz}^{\text{exp}}$ | theoretical/experimental longitudinal nominal stress |
| $e$ | engineering strain |
| $m$ | (machine precision)/2 |
| $\delta, \Delta$ | relative/absolute error |
| $\bar{\delta}, \bar{\Delta}$ | average relative/absolute error |
| $\delta_{\max}, \Delta_{\max}$ | maximum relative/absolute error |
| $x, y, z$ | Cartesian coordinates in deformed configuration |
| $X, Y, Z$ | Cartesian coordinates in undeformed configuration |
| $\gamma_1, \gamma_2$ | shear strains |
| $T_1, T_2$ | shear stresses |
| $\chi$ | function defined in Eq. (64) |
| $C$ | constant of integration defined in Eq. (68) |
| $N$ | resultant axial load acting on tendon |
| $S$ | average force per unit undeformed area acting on tendon |

Two examples of microstructural SEFs are those derived by Grytz and Meschke (2009) and Shearer (2015). Both are based on the geometrical arrangement of the fibrils within the fascicle and neglect any subunits below the fibril level (Fig. 2). Shearer considered a fascicle whose fibrils are coaligned with its axis, but have a distribution of crimp levels throughout its radius (Fig. 3 (1) and (2)), whereas Grytz and Meschke considered a helical arrangement of fibrils, but neglected their crimp (Fig. 3(3) and (4)). Grytz and Meschke defined the angle that these fibrils make with the fascicle's longitudinal axis as the crimp angle, but this is not the usual definition of crimp. Here, this quantity is referred to as the *fibril helix angle*. The logical extension of these models is to allow the fibrils to be helically arranged *and* crimped (Fig. 3(5)); this is the case considered here. A scanning electron microscope (SEM) image displaying fibrils that are both helically arranged *and* crimped appears in Fig. 9 of Yahia and Drouin (1989).

A considerable amount of work has been dedicated to modelling other types of fibre-reinforced composite materials. Crossley et al. (2003), for example, derived analytical solutions that govern the bending and flexure of helically reinforced, anisotropic, linear elastic cylinders. This is built on a large body of literature on modelling cables (Cardou and Jolicouer, 1997) and rope (Costello, 1978, 1997). Adkins and Rivlin (1955) discussed finite deformations of materials that are reinforced by inextensible cords, and Spencer and Soldatos (2007) considered finite deformations of fibre-reinforced elastic solids whose fibres are capable of resisting bending. Whilst these general theories are extremely valuable for certain problems, to model the behaviour of a material with a microstructure as complex as a ligament or tendon requires a more specific model.

In this paper, a new SEF that governs the behaviour of a ligament or tendon with the microstructure described above is derived. In Section 2, the stress–strain response of a single fascicle is calculated and this relationship is used to determine the form of the new SEF in Section 3. The SEF is used to model the mechanical response of human patellar tendon to uniaxial extension and shear in Section 4. In Section 5, the case of a ligament or tendon with helical *fascicles* is explored and the relative effects of the *fibril crimp angle*, *fibril helix angle* and *fascicle helix angle* are analysed. Finally, the implications of the model are discussed in Section 6.

## 2. The stress–strain response of a fascicle with helically aligned fibrils

Kastelic et al. (1980) derived the stress–strain response of a fascicle with fibrils that are coaligned with its longitudinal axis, and Shearer (2015) adapted their method to derive analytical expressions for these relationships for different fibril crimp angle distributions. Here, this work is extended to ligaments and tendons whose fascicles have a helical arrangement of fibrils.

Kastelic et al. (1978) observed that crimp angle varies throughout the radius of a fascicle with longitudinal fibrils. It was then noted by Yahia and Drouin (1989) that this is also the case in fascicles with a helical arrangement of fibrils. The minimum crimp angle occurs at the fascicle's centre, the maximum at its edge. Therefore, assuming that only fully extended fibrils contribute to the fascicle's resistance to an applied load, its rigidity will increase as more and more fibrils become fully extended. The fibrils at the centre of the fascicle are the first to tauten, then as the fascicle is increasingly stretched, more and more fibrils within a circle of increasing radius begin to contribute to its stiffness until the outer fibrils are finally fully extended. At this point, the toe-region ends and the linear region begins.

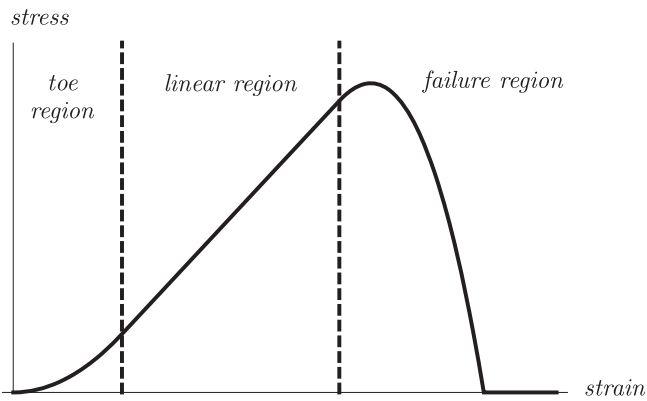

**Fig. 1.** Schematic representation of typical ligament and tendon stress-strain behaviour.

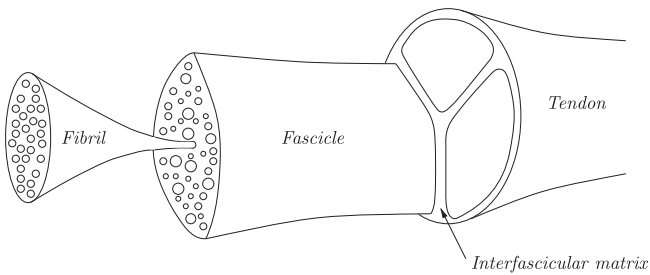

**Fig. 2.** Idealised tendon hierarchy (adapted from Shearer, 2015) used by Grytz and Meschke (2009) and Shearer (2015).

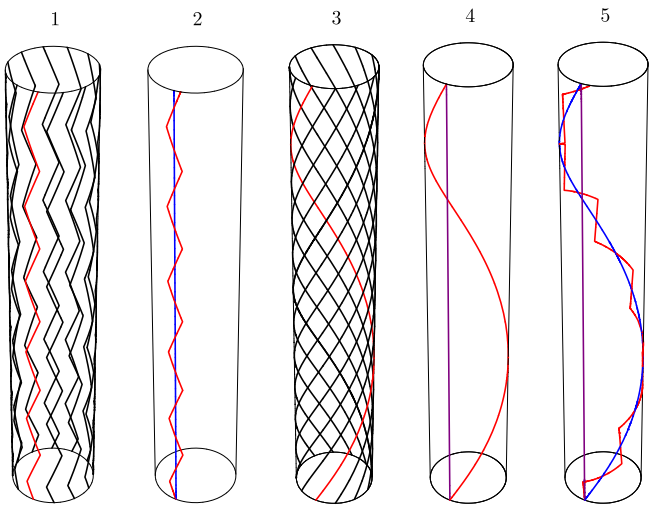

**Fig. 3.** Schematic diagrams of fascicles: (1) crimped fibrils coaligned with fascicle longitudinal axis, as modelled by Shearer (2015), (2) the angle the crimped fibril (red) makes with the blue line is the *fibril crimp angle*, (3) uncrimped helical fibrils, as modelled by Grytz and Meschke (2009), (4) the angle the fibril (red) makes with the purple line is the *fibril helix angle*, (5) crimped, helical fibril, as modelled in the present paper; the angle the crimped fibril (red) makes with the blue line is the *fibril crimp angle*, the angle the blue line makes with the purple line is the *fibril helix angle*. (For interpretation of the references to color in this figure caption, the reader is referred to the web version of this paper.)

It is assumed that the crimp angle distribution is described by Shearer (2015)

$$\theta_p(\rho) = \sin^{-1}(\sin(\theta_o)\rho^p), \quad (1)$$

where $\rho$ is a non-dimensional radial variable scaled on the fascicle's radius so that $0 \leq \rho \leq 1$, $\theta_o$ is the crimp angle of the outer fibrils and $p$ is a parameter that can be chosen to fit an observed crimp distribution. It is assumed that the fibrils make an angle $\alpha$ with the fascicle's longitudinal axis, so that we can write a unit vector pointing in the fibril direction in cylindrical coordinates as follows:

$$\mathbf{P} = (0, \sin\alpha, \cos\alpha). \quad (2)$$

In the following, it is important to distinguish between the fascicle stretch $\lambda$, the component of that stretch in the fibril direction $\Lambda$, and the stretch actually experienced by a fibril, which is only greater than one once its crimp has straightened out. If the fascicle undergoes a longitudinal stretch $\lambda$ and we assume that fibrils slide freely without interacting with each other, or with the extra-collagenous matrix that holds them in place, then the fibril alignment vector will be modified to

$$\mathbf{p} = (0, \sin\alpha, \lambda\cos\alpha). \quad (3)$$

In reality, the fibrils *are* likely to interact with each other and fibril sliding is likely to affect ligament and tendon viscoelasticity (Screen, 2008); however, to keep the model as simple as possible, it is assumed that fibril interaction is negligible in the elastic case being considered here. The component of the fascicle stretch in the fibril direction is

$$\Lambda = \frac{|\mathbf{p}|}{|\mathbf{P}|} = \sqrt{\sin^2\alpha + \lambda^2\cos^2\alpha}. \quad (4)$$

This stretch straightens out the fibrils' crimp until they become taut. At this point, the fibrils themselves begin to stretch. The stretch in the fibril direction required to straighten the fibrils at radius $\rho$ is

$$\Lambda_p(\rho) = \frac{1}{\cos(\theta_p(\rho))} = \frac{1}{\sqrt{1 - \sin^2(\theta_o)\rho^{2p}}}, \quad (5)$$

and that required to straighten the outer fibrils is

$$\Lambda^* = \frac{1}{\cos\theta_o}. \quad (6)$$

Using Eq. (4), the fascicle stretch corresponding to $\Lambda^*$ can be calculated

$$\lambda^* = \frac{1}{\cos\alpha}\sqrt{\frac{1}{\cos^2\theta_o} - \sin^2\alpha}. \quad (7)$$

At the critical fascicle strain $\epsilon^* = \lambda^* - 1$, the toe-region ends. For a fascicle strain $\epsilon = \lambda - 1$ satisfying $0 \leq \epsilon \leq \epsilon^*$, however, there is an internal area of taut fibrils, each carrying a share of the load. The radius $R_p$ of this circle can be determined from (5), by equating $\Lambda = \Lambda_p(R_p)$

$$\theta_p(R_p) = \cos^{-1}\left(\frac{1}{\Lambda}\right). \quad (8)$$

Upon using (1) in (8), we obtain

$$R_p = \left(\frac{1}{\sin^2\theta_o}\left(1 - \frac{1}{\Lambda^2}\right)\right)^{1/2p}. \quad (9)$$

Fibrils outside $R_p$ retain their crimping and experience no load. As the fascicle is stretched, this radius increases until $R_p = 1$, at which point all fibrils are finally taut.

The tensile load experienced by the fascicle $P_p$ is

$$P_p = \int_0^{R_p} \sigma_p(\rho) 2\pi\rho \, d\rho, \tag{10}$$

where $\sigma_p(\rho)$ is the contribution of the Cauchy stress in the fibrils at radius $\rho$ in the fascicle direction, and the upper limit of integration is determined by (9).

The stress $\sigma_p(\rho)$ is related to the fibril stress $\sigma_p^f(\rho)$ by the following expression:

$$\sigma_p(\rho) = \sigma_p^f(\rho) \frac{(0, \sin\alpha, \lambda\cos\alpha)\cdot(0,0,1)}{\sqrt{\sin^2\alpha + \lambda^2\cos^2\alpha}}$$
$$= \sigma_p^f(\rho) \frac{\lambda\cos\alpha}{\sqrt{\sin^2\alpha + \lambda^2\cos^2\alpha}}, \tag{11}$$

therefore

$$P_p = \frac{\lambda\cos\alpha}{\sqrt{\sin^2\alpha + \lambda^2\cos^2\alpha}} \int_0^{R_p} \sigma_p^f(\rho) 2\pi\rho \, d\rho. \tag{12}$$

We assume that the fibrils obey a linear stress–strain relationship (as observed by Sasaki and Odajima, 1996)

$$\sigma_p^f(\rho) = E\epsilon_p^f(\rho), \tag{13}$$

where $E$ is the fibril Young's modulus, and the fibril strain $\epsilon_p^f(\rho)$ for a given fibril within the radius $R_p$ is (Shearer, 2015)

$$\epsilon_p^f(\rho) = \Lambda \cos(\theta_p(\rho)) - 1. \tag{14}$$

Upon substituting (13) and (14) into (12), an expression for the average traction in the fascicle direction is derived

$$\tau_p = \frac{P_p}{\pi} = 2E\lambda \cos\alpha \int_0^{R_p} \left( \cos(\theta_p(\rho)) - \frac{1}{\sqrt{\sin^2\alpha + \lambda^2\cos^2\alpha}} \right) \rho \, d\rho$$
$$= 2E\lambda \cos\alpha \int_0^{R_p} \left( \sqrt{1 - \rho^{2p}\sin^2\theta_o} - \frac{1}{\sqrt{\sin^2\alpha + \lambda^2\cos^2\alpha}} \right) \rho \, d\rho. \tag{15}$$

For certain values of $p$, (15) can be evaluated explicitly. For $p=1,2$, we obtain

$$\tau_1 = \frac{E\lambda \cos\alpha}{3\sin^2\theta_o} \left( 2 - \frac{3}{\sqrt{\sin^2\alpha + \lambda^2\cos^2\alpha}} + \frac{1}{(\sin^2\alpha + \lambda^2\cos^2\alpha)^{3/2}} \right), \tag{16}$$

$$\tau_2 = \frac{E\lambda \cos\alpha}{2\sin^2\theta_o} \left( \frac{1}{\sin^{-1}\left(\Lambda/\sqrt{(\Lambda+1)(\Lambda-1)}\right)} - \frac{\sqrt{(\Lambda+1)(\Lambda-1)}}{\Lambda^2} \right). \tag{17}$$

The simpler expression is $\tau_1$, so this is used to derive the SEF in the following section. Eq. (16) only holds for $0 \leq \lambda \leq \lambda_*$; for $\lambda > \lambda_*$

$$\tau_1 = 2E\lambda \cos\alpha \int_0^1 \left( \sqrt{1 - \rho^2 \sin^2\theta_o} - \frac{1}{\sqrt{\sin^2\alpha + \lambda^2\cos^2\alpha}} \right) \rho \, d\rho$$
$$= E\lambda \cos\alpha \left( \beta - \frac{1}{\sqrt{\sin^2\alpha + \lambda^2\cos^2\alpha}} \right), \quad \beta = \frac{2(1 - \cos^3\theta_o)}{3\sin^2\theta_o}. \tag{18}$$

The critical stretch $\lambda_*$ at which the toe-region ends is a monotonically increasing function of $\alpha$ and $\theta_o$, provided $0 \leq \alpha \leq \pi/2$, $0 \leq \theta_o \leq \pi/2$; however, if $\theta_o = 0$, then $\lambda_* = 1$ and there is no toe-region. This makes sense since, if $\theta_o = 0$, all the fibrils have no crimp and contribute to the tendon's stiffness from the start of its deformation.

## 3. Derivation of the strain energy function

The ligament or tendon under consideration is modelled as incompressible, anisotropic and hyperelastic and is characterised via the SEF $W$. It is assumed that $W$ is a function of the strain invariants $I_1$ and $I_4$ only

$$W(I_1, I_4) = (1 - \phi)W_m(I_1) + \phi W_f(I_4), \tag{19}$$

where $\phi$ is the collagen volume fraction. In the current context, $W_f$ is the strain energy associated with the fascicles and $W_m$ is that associated with the extracollagenous matrix. The invariants $I_1$ and $I_4$ are defined by

$$I_1 = \text{tr } \mathbf{C}, \quad I_4 = \mathbf{M}\cdot(\mathbf{CM}), \tag{20}$$

where $\mathbf{C} = \mathbf{F}^T\mathbf{F}$ is the right Cauchy–Green tensor ($\mathbf{F}$ is the deformation gradient, Ogden, 1997), and $\mathbf{M}$ is a unit vector pointing in fascicle direction in the undeformed configuration. $I_4$ can be interpreted as the square of the stretch in the fascicle direction

$$I_4 = M_i C_{ij} M_j = M_i F_{ki} F_{kj} M_j = (\mathbf{FM})\cdot(\mathbf{FM}) = \mathbf{m}\cdot\mathbf{m} = |\mathbf{m}|^2, \tag{21}$$

where $\mathbf{m} = \mathbf{FM}$ is the push forward of $\mathbf{M}$ to the deformed configuration.

Given the SEF above, the Cauchy stress tensor takes the following form (this can be seen by taking $\partial W/\partial I_2 = \partial W/\partial I_5 = 0$ in Eq. (2.6) of Holzapfel and Ogden, 2010)

$$\mathbf{T} = -Q\mathbf{I} + 2W_1\mathbf{B} + 2W_4 \mathbf{m} \otimes \mathbf{m}, \tag{22}$$

where $Q$ is a Lagrange multiplier associated with the incompressibility constraint, $\mathbf{I}$ is the identity tensor, $W_i = \partial W/\partial I_i$, and $\mathbf{B} = \mathbf{FF}^T$ is the left Cauchy–Green tensor. $\mathbf{T}_f = 2W_4 \mathbf{m} \otimes \mathbf{m}/\phi$ is the component of the Cauchy stress associated with the fascicles, which will be used to derive an expression equivalent to $\tau_1$ in (16) and (18).

The traction associated with $\mathbf{T}_f$ acting on a face normal to the deformed fascicle direction (with unit normal $\hat{\mathbf{m}} = \mathbf{m}/|\mathbf{m}|$) is

$$\mathbf{t}_f = \mathbf{T}_f\cdot\left(\frac{\mathbf{m}}{|\mathbf{m}|}\right) = 2\frac{W_4}{\phi}(\mathbf{m} \otimes \mathbf{m})\cdot\left(\frac{\mathbf{m}}{|\mathbf{m}|}\right) = 2\frac{W_4}{\phi}\mathbf{m}\frac{\mathbf{m}\cdot\mathbf{m}}{|\mathbf{m}|} = 2\frac{W_4}{\phi}|\mathbf{m}|\mathbf{m}. \tag{23}$$

The component of this traction in the fascicle direction is

$$\frac{\mathbf{t}_f\cdot\mathbf{m}}{|\mathbf{m}|} = \frac{(2W_4|\mathbf{m}|\mathbf{m})\cdot\mathbf{m}}{\phi|\mathbf{m}|} = 2\frac{W_4}{\phi}|\mathbf{m}|^2 = 2\frac{W_4}{\phi}I_4. \tag{24}$$

By equating (24) with (16) and (18), two equations for the required form of $W_f$ are obtained

$$2I_4 \frac{dW_f}{dI_4} = \frac{E\lambda \cos\alpha}{3\sin^2\theta_o} \left( 2 - \frac{3}{\sqrt{\sin^2\alpha + \lambda^2\cos^2\alpha}} + \frac{1}{(\sin^2\alpha + \lambda^2\cos^2\alpha)^{3/2}} \right), \quad 1 \leq \lambda \leq \lambda_*, \tag{25}$$

$$2I_4 \frac{dW_f}{dI_4} = E\lambda \cos\alpha \left( \beta - \frac{1}{\sqrt{\sin^2\alpha + \lambda^2\cos^2\alpha}} \right), \quad \lambda > \lambda_*. \tag{26}$$

Since $I_4$ is the square of the stretch in the fascicle direction, $\lambda = \sqrt{I_4}$; hence

$$\frac{dW_f}{dI_4} = \frac{E\cos\alpha}{6\sqrt{I_4}\sin^2\theta_o} \left( 2 - \frac{3}{\sqrt{\sin^2\alpha + I_4\cos^2\alpha}} + \frac{1}{(\sin^2\alpha + I_4\cos^2\alpha)^{3/2}} \right), \quad 1 \leq I_4 \leq \lambda_*^2, \tag{27}$$

$$\frac{dW_f}{dI_4} = \frac{E \cos \alpha}{2\sqrt{I_4}} \left( \beta - \frac{1}{\sqrt{\sin^2 \alpha + I_4 \cos^2 \alpha}} \right), \quad I_4 > \lambda^{*2}, \tag{28}$$

Eqs. (27) and (28) can be integrated to find the required form for the anisotropic component of the SEF

$$W_f = \frac{E}{3 \sin^2 \theta_o}$$
$$\left( 2 \cos \alpha \sqrt{I_4} - 3 \log \left( \cos^2 \alpha \sqrt{I_4} + \cos \alpha \sqrt{\sin^2 \alpha + I_4 \cos^2 \alpha} \right) \right.$$
$$\left. + \frac{\cos \alpha \sqrt{I_4}}{\sin^2 \alpha \sqrt{\sin^2 \alpha + I_4 \cos^2 \alpha}} \right) + \gamma, \quad 1 \leq I_4 \leq \lambda^{*2}, \tag{29}$$

$$W_f = E \left( \beta \cos \alpha \sqrt{I_4} - \log \left( \cos^2 \alpha \sqrt{I_4} + \cos \alpha \sqrt{\sin^2 \alpha + I_4 \cos^2 \alpha} \right) \right) + \eta, \quad I_4 > \lambda^{*2} \tag{30}$$

where $\gamma$ and $\eta$ are constants of integration, which must be chosen to satisfy $W_f|_{I_4=1} = 0$ and ensure that $W_f$ is continuous at

$$I_4 = \lambda^{*2} = \frac{1}{\cos^2 \alpha} \left( \frac{1}{\cos^2 \theta_o} - \sin^2 \alpha \right). \tag{31}$$

Upon applying these conditions, we find

$$\gamma = -\frac{E}{3 \sin^2 \theta_o} \left( 2 \cos \alpha - 3 \log \left( \cos^2 \alpha + \cos \alpha \right) + \frac{\cos \alpha}{\sin^2 \alpha} \right), \tag{32}$$

$$\eta = \gamma + E \left( \left( \frac{\cos \theta_o}{\sin^2 \theta_o} \frac{1}{\sin^2 \alpha} + \frac{2}{\sin^2 \alpha} - 3\beta \right) \sqrt{\frac{1}{\cos^2 \theta_o} - \sin^2 \alpha} \right.$$
$$\left. -3 \frac{\cos^2 \theta_o}{\sin^2 \theta_o} \log \left( \cos \alpha \left( \frac{1}{\cos \theta_o} + \sqrt{\frac{1}{\cos^2 \theta_o} - \sin^2 \alpha} \right) \right) \right). \tag{33}$$

Finally, note that for $I_4 < 1$, $W_f = 0$. Therefore, the anisotropic part of the SEF is now explicit.

As in Shearer (2015), the isotropic component of the SEF is chosen to be neo-Hookean

$$W_m(I_1) = \frac{\mu}{2}(I_1 - 3), \tag{34}$$

where $\mu > 0$ is the ground state shear modulus of the extracollagenous matrix. The neo-Hookean SEF accurately models the behaviour of arterial elastin (Gundiah et al., 2007) and it is assumed that the extracollagenous matrix in ligaments in tendons has similar mechanical properties.

The complete form of the SEF can now be written explicitly

$$W = (1 - \phi) \frac{\mu}{2} (I_1 - 3), \quad I_4 < 1, \tag{35}$$

$$W = (1 - \phi) \frac{\mu}{2} (I_1 - 3) + \phi \frac{E}{3 \sin^2 \theta_o} \left( 2 \cos \alpha \sqrt{I_4} \right.$$
$$\left. - 3 \log \left( 2 \left( \cos^2 \alpha \sqrt{I_4} + \cos \alpha \sqrt{\sin^2 \alpha + I_4 \cos^2 \alpha} \right) \right) + \frac{\cos \alpha \sqrt{I_4}}{\sin^2 \alpha \sqrt{\sin^2 \alpha + I_4 \cos^2 \alpha}} \right) + \gamma, \quad 1 \leq I_4 \leq \lambda^{*2}, \tag{36}$$

$$W = (1 - \phi) \frac{\mu}{2} (I_1 - 3) + \phi E \left( \beta \cos \alpha \sqrt{I_4} \right.$$
$$\left. - \log \left( \cos^2 \alpha \sqrt{I_4} + \cos \alpha \sqrt{\sin^2 \alpha + I_4 \cos^2 \alpha} \right) \right) + \eta, \quad I_4 > \lambda^{*2}. \tag{37}$$

The parameters $\lambda_*, \beta, \gamma$ and $\eta$ are defined in Eqs. (7), (18), (32) and (33), respectively. Note that by taking the limit as $\alpha \to 0$ in the above, we obtain

$$W = (1 - \phi) \frac{\mu}{2} (I_1 - 3), \quad I_4 < 1, \tag{38}$$

$$W = (1 - \phi) \frac{\mu}{2} (I_1 - 3) + \frac{\phi E}{6 \sin^2 \theta_o} \left( 4 \sqrt{I_4} - 3 \log(I_4) - \frac{1}{\sqrt{I_4} - 3} \right),$$
$$1 \leq I_4 \leq \frac{1}{\cos^2 \theta_o}, \tag{39}$$

$$W = (1 - \phi) \frac{\mu}{2} (I_1 - 3) + \phi E \left( \beta \sqrt{I_4} - \frac{1}{2} \log(I_4) + \hat{\eta} \right),$$
$$I_4 > \frac{1}{\cos^2 \theta_o}, \tag{40}$$

where

$$\hat{\eta} = -\frac{1}{2} - \frac{\cos^2 \theta_o}{\sin^2 \theta_o} \log \left( \frac{1}{\cos \theta_o} \right). \tag{41}$$

This is the SEF for a ligament or tendon whose fascicles have longitudinal fibrils (Shearer, 2015).

## 4. Human patellar tendon

### 4.1. Unixial extension

The SEF derived in the previous section is now used to determine a stress–strain relationship, which is compared with the experimental data for younger human patellar tendons from (Johnson et al., 1994). For simplicity, the tendon is modelled as a circular cylinder.

We assume that the tendon has radius $A$ and length $L$ in the undeformed configuration. Its geometry is described in terms of the cylindrical coordinates $(R, \Theta, Z)$, so $0 \leq R \leq A$, $0 \leq \Theta < 2\pi$, $0 \leq Z \leq L$. After applying a homogeneous longitudinal stretch, the corresponding deformed coordinates are $(r, \theta, z)$, so $0 \leq r \leq a$, $0 \leq \theta < 2\pi$, $0 \leq z \leq l$, where $a$ and $l$ are the deformed counterparts of $A$ and $L$. The deformation, then, is described by

$$r = \frac{R}{\sqrt{\zeta}}, \quad \theta = \Theta, \quad z = \zeta Z, \tag{42}$$

where $\zeta = l/L$ is the longitudinal stretch, and the first equation is a consequence of the tendon's incompressibility. The deformation gradient is

$$\mathbf{F} = F_{iJ} \mathbf{e}_i \otimes \mathbf{E}_J, \quad F_{iJ} = \begin{pmatrix} \zeta^{-1/2} & 0 & 0 \\ 0 & \zeta^{-1/2} & 0 \\ 0 & 0 & \zeta \end{pmatrix}, \tag{43}$$

where $\mathbf{e}_i$, $i = (r, \theta, z)$, and $\mathbf{E}_J$, $J = (R, \Theta, Z)$, are deformed and undeformed unit vectors in the radial, azimuthal, and longitudinal directions, respectively. The left Cauchy–Green tensor is

$$\mathbf{B} = B_{ij} \mathbf{e}_i \otimes \mathbf{e}_j, \quad B_{ij} = \begin{pmatrix} \zeta^{-1} & 0 & 0 \\ 0 & \zeta^{-1} & 0 \\ 0 & 0 & \zeta^2 \end{pmatrix}. \tag{44}$$

Since the patellar tendon's fascicles are coaligned with its longitudinal axis, $\mathbf{M} = \mathbf{E}_Z$, therefore, $\mathbf{m} = \zeta \mathbf{e}_z$. $I_1$ and $I_4$ are given by

$$I_1 = \frac{2}{\zeta} + \zeta^2, \quad I_4 = \zeta^2. \tag{45}$$

The Cauchy stress tensor is

$$\mathbf{T} = T_{ij}\mathbf{e}_i \otimes \mathbf{e}_j, \quad T_{ij} = \begin{pmatrix} T_{rr} & 0 & 0 \\ 0 & T_{rr} & 0 \\ 0 & 0 & T_{zz} \end{pmatrix}, \quad (46)$$

where

$$T_{rr} = (1-\phi)\mu\zeta^{-1} - Q, \quad (47)$$

$$T_{zz} = \begin{cases} (1-\phi)\mu\zeta^2 - Q, & \zeta < 1, \\ (1-\phi)\mu\zeta^2 + \phi\dfrac{E\zeta\cos\alpha}{3\sin^2\theta_o}\left(2 - \dfrac{3}{\sqrt{\sin^2\alpha + \zeta^2\cos^2\alpha}} \right. \\ \left. + \dfrac{1}{(\sin^2\alpha + \zeta^2\cos^2\alpha)^{3/2}}\right) - Q, & 1 \leq \zeta \leq \lambda^*, \\ (1-\phi)\mu\zeta^2 + \phi E\zeta\cos\alpha\left(\beta - \dfrac{1}{\sqrt{\sin^2\alpha + \zeta^2\cos^2\alpha}}\right) - Q, & \zeta > \lambda^*. \end{cases} \quad (48)$$

The static equations of equilibrium in the absence of body forces are

$$\text{div } \mathbf{T} = \mathbf{0}, \quad (49)$$

where div is the divergence operator in the deformed configuration. In this case, (49) reduces to

$$\frac{\partial Q}{\partial r} = \frac{\partial Q}{\partial \theta} = \frac{\partial Q}{\partial z} = 0. \quad (50)$$

Therefore, $Q$ is a constant, which can be determined by applying a traction-free boundary condition on $r=a$ ($\mathbf{T}\cdot\mathbf{n} = \mathbf{0}$, where $\mathbf{n} = \mathbf{e}_r$), which gives $T_{rr}|_{r=a} = T_{r\theta}|_{r=a} = T_{rz}|_{r=a} = 0$. The first of these gives

$$Q = (1-\phi)\mu\zeta^{-1}; \quad (51)$$

hence $T_{rr} = 0$, and an explicit expression for $T_{zz}$ is obtained

$$T_{zz} = \begin{cases} (1-\phi)\mu(\zeta^2 - \zeta^{-1}), & \zeta < 1, \\ (1-\phi)\mu(\zeta^2 - \zeta^{-1}) + \phi\dfrac{E\zeta\cos\alpha}{3\sin^2\theta_o} \times \\ \left(2 - \dfrac{3}{\sqrt{\sin^2\alpha + \zeta^2\cos^2\alpha}} \right. \\ \left. + \dfrac{1}{(\sin^2\alpha + \zeta^2\cos^2\alpha)^{3/2}}\right), & 1 \leq \zeta \leq \lambda^*, \\ (1-\phi)\mu(\zeta^2 - \zeta^{-1}) \\ + \phi E\zeta\cos\alpha\left(\beta - \dfrac{1}{\sqrt{\sin^2\alpha + \zeta^2\cos^2\alpha}}\right), & \zeta > \lambda^*. \end{cases} \quad (52)$$

The corresponding nominal stress, which gives the force per unit *undeformed* area (the correct measure of stress to use when comparing with experimental data), is given by $S_{zz} = T_{zz}/\zeta$. In Shearer (2015), $S_{zz}$ (but with $\alpha=0°$, since the patellar tendon's fascicles were assumed to have longitudinal fibrils) was fitted to the data from Johnson et al. (1994). Since the stiffness of a ligament's or tendon's matrix is negligible compared to that of its fascicles, $(1-\phi)\mu$ was chosen to be small ($(1-\phi)\mu = 0.01$ MPa) relative to the reported values of $E$, which range from 32 MPa (Graham et al., 2004) to 11 GPa (Wenger et al., 2007). The remaining parameters were determined using the FindFit function in *Mathematica 7* (Wolfram Research, Inc., Champaign, IL, 2008) subject to the conditions

$$\phi E > 0, \quad 0 \leq \theta_o \leq \pi/2. \quad (53)$$

The values determined by FindFit were $\phi E = 558$ MPa, and $\theta_o = 0.19$ rad $= 10.7°$. FindFit applies a least-squares method to find a locally optimal solution in the non-linear case (which is the case here); therefore, the restrictions (53) are crucial to ensure that the given solution is physically realistic. Using the default settings, it attempts to make the numerical error in a result of size $x$ be less than $10^{-m} + |x|10^{-m}$, where $m = $ (machine precision)/2.

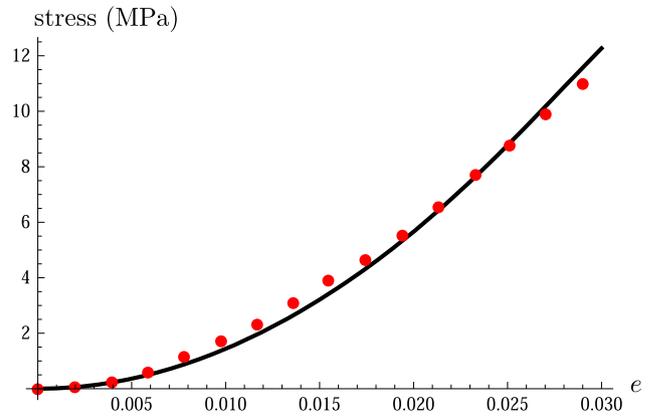

**Fig. 4.** Stress–strain curve showing the ability of the model to reproduce experimental data taken from Johnson et al. (1994). Solid black: model, red circles: experimental data. Parameter values: $(1-\phi)\mu = 0.01$ MPa, $\phi E = 1027$ MPa, $\theta_o = 0.17$ rad $= 9.5°$, $\alpha = 27°$. (For interpretation of the references to color in this figure caption, the reader is referred to the web version of this paper.)

A crude measurement of the helix angle shown in the SEM image in Fig. 9 of Yahia and Drouin (1989) using a protractor gives $\alpha \approx 27°$ (this image is of a canine patellar tendon fascicle, but it is assumed that human patellar tendon fascicles have a similar fibril helix angle). Using this value, and again choosing $(1-\phi)\mu = 0.01$ MPa, FindFit was again used to predict the values of $\phi E$ and $\theta_o$ subject to the same conditions (53). To ensure that realistic results were obtained, the values of $\phi E$ and $\theta_o$ derived when $\alpha = 0°$ were used as initial "guesses". The values determined this time were $\phi E = 1027$ MPa and $\theta_o = 0.20$ rad $= 11.5°$. A comparison between the experimental and theoretical stresses, using this set of parameters, as a function of the engineering strain $e = \zeta - 1$, is plotted in Fig. 4. The predicted values of $\phi E$ and $\theta_o$ are highly dependent on the value of $\alpha$; therefore, this parameter cannot, in general, be neglected.

The effectiveness of the new model can be quantified in terms of the relative and absolute errors, defined respectively by

$$\delta = |S_{zz}^{exp} - S_{zz}|/|S_{zz}^{exp}|, \quad \Delta = |S_{zz}^{exp} - S_{zz}|, \quad (54)$$

where $S_{zz}^{exp}$ is the experimental stress. The average values of these quantities were $\bar{\delta} = 9.8\%$ and $\bar{\Delta} = 0.24$ MPa. Their maxima were $\delta_{max} = 24.8\%$ and $\Delta_{max} = 0.57$ MPa.

### 4.2. Shear

Whilst *ex vivo* uniaxial extension data collected in the laboratory provides useful information about the mechanical behaviour of a ligament or tendon, *in vivo* loading conditions are more complex and are likely to involve a combination of both extension and *shear*. Shear tests can be extremely challenging to undertake due to the difficulties involved in gripping samples. Additionally, to obtain a cubic sample suitable for a shear test, the specimen must be dissected, which could potentially affect its mechanical properties due to damage to its ultrastructure. Mathematical modelling allows us to simulate a shear test, whilst avoiding any experimental difficulties.

Here, two distinct modes of simple shear are considered: those in which planes parallel to the fascicle direction remain undeformed:

$$\mathbf{F} = F_{ij}\mathbf{e}_i \otimes \mathbf{E}_J, \quad F_{ij} = \begin{pmatrix} 1 & \gamma_1 & 0 \\ 0 & 1 & 0 \\ 0 & 0 & 1 \end{pmatrix}, \quad F_{ij} = \begin{pmatrix} 1 & 0 & 0 \\ \gamma_1 & 1 & 0 \\ 0 & 0 & 1 \end{pmatrix}, \tag{55}$$

$$\mathbf{F} = F_{ij}\mathbf{e}_i \otimes \mathbf{E}_J, \quad F_{ij} = \begin{pmatrix} 1 & 0 & 0 \\ 0 & 1 & 0 \\ \gamma_1 & 0 & 1 \end{pmatrix}, \quad F_{ij} = \begin{pmatrix} 1 & 0 & 0 \\ 0 & 1 & 0 \\ 0 & \gamma_1 & 1 \end{pmatrix}, \tag{56}$$

and those in which planes perpendicular to it remain undeformed

$$\mathbf{F} = F_{ij}\mathbf{e}_i \otimes \mathbf{E}_J, \quad F_{ij} = \begin{pmatrix} 1 & 0 & \gamma_2 \\ 0 & 1 & 0 \\ 0 & 0 & 1 \end{pmatrix}, \quad F_{ij} = \begin{pmatrix} 1 & 0 & 0 \\ 0 & 1 & \gamma_2 \\ 0 & 0 & 1 \end{pmatrix}, \tag{57}$$

where $i = (x, y, z)$, $J = (X, Y, Z)$, $\gamma_1$ and $\gamma_2$ are the shear strains and it is assumed that the fascicles are initially aligned in the $Z$-direction so that $\mathbf{M} = \mathbf{E}_Z$. Upon substituting (55)–(57) into (22), the corresponding shear stresses are obtained

$$T_1 = (1 - \phi)\mu\gamma_1, \tag{58}$$

$$T_2 = \begin{cases} (1 - \phi)\mu\gamma_2 + \phi\dfrac{E\gamma_2 \cos\alpha}{3\sqrt{1 + \gamma_2^2 \sin^2\theta_o}} \times \\ \quad \left(2 - \dfrac{3}{\sqrt{1 + \gamma_2^2 \cos^2\alpha}} \right. & \gamma_2^2 \leq \lambda^{*2} - 1, \\ \quad \left. + \dfrac{1}{(1 + \gamma_2^2 \cos^2\alpha)^{3/2}}\right), \\ (1 - \phi)\mu\gamma_2 & \gamma_2^2 > \lambda^{*2} - 1. \\ \quad + \phi\dfrac{E\gamma_2 \cos\alpha}{\sqrt{1 + \gamma_2^2}}\left(\beta - \dfrac{1}{\sqrt{1 + \gamma_2^2 \cos^2\alpha}}\right), \end{cases} \tag{59}$$

Using the parameter values derived for human patellar tendon, above, these shear stresses are plotted in Fig. 5. As can be seen, the model predicts that:

- the response of the patellar tendon to shear in which planes parallel to the fascicle length remain undeformed is linear,
- it experiences strain stiffening under shear in which planes perpendicular to the fascicle direction remain undeformed,

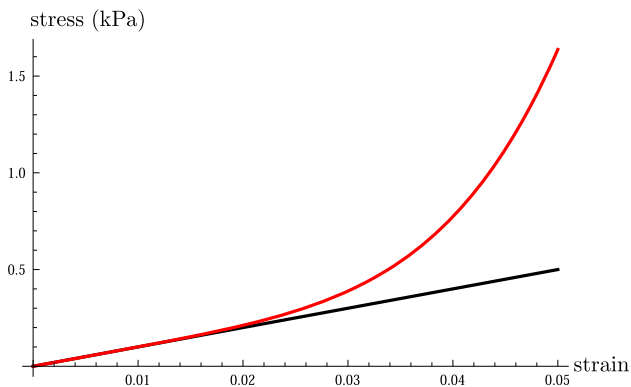

**Fig. 5.** Stress–strain curves showing the theoretical response of human patellar tendon to shear. Black: $T_1$ as a function of $\gamma_1$, red: $T_2$ as a function of $\gamma_2$. Parameter values: $(1 - \phi)\mu = 0.01$ MPa, $\phi E = 1027$ MPa, $\theta_o = 0.17$ rad $= 9.5°$, $\alpha = 27°$. (For interpretation of the references to color in this figure caption, the reader is referred to the web version of this paper.)

- in both cases, the stresses required to shear the tendon are much smaller than those required to uniaxially stretch it to the same level of strain.

## 5. Ligaments and tendons with helical fascicles

In addition to having fascicles with helically arranged fibrils, some ligaments and tendons such as the anterior cruciate ligament (Shearer et al., 2014) and the extensor carpi ulnaris tendon (Kalson et al., 2011) also have *fascicles that form a helix* with their longitudinal axis. To model this, the fascicle alignment vector $\mathbf{M}$ must be modified so that it no longer only points in the $Z$-direction, but also has a component in the $\Theta$-direction

$$\mathbf{M} = \sin\psi\mathbf{E}_\Theta + \cos\psi\mathbf{E}_Z, \tag{60}$$

where $\psi$ is the angle that the fascicles make with the ligament's or tendon's longitudinal axis. Upon doing this, and again assuming the ligament or tendon undergoes a stretch $\zeta$, we obtain a modified Cauchy stress tensor

$$\mathbf{T} = T_{ij}\mathbf{e}_i \otimes \mathbf{e}_j, \quad T_{ij} = \begin{pmatrix} T_{rr} & 0 & 0 \\ 0 & T_{\theta\theta} & T_{\theta z} \\ 0 & T_{\theta z} & T_{zz} \end{pmatrix}, \tag{61}$$

where $T_{rr} = T_{rr} = (1 - \phi)\mu\zeta^{-1} - Q$, as before, and

$$T_{\theta\theta} = (1 - \phi)\mu\zeta^{-1} + 2\zeta^{-1}\chi \sin^2\psi - Q, \quad T_{\theta z} = 2\zeta^{1/2}\chi \cos\psi \sin\psi, \tag{62}$$

$$T_{zz} = (1 - \phi)\mu\zeta^2 + 2\zeta^2\chi \cos^2\psi - Q, \tag{63}$$

where

$$\chi = \begin{cases} 0, & I_4 < 1 \\ \phi\dfrac{E\cos\alpha}{6\sqrt{I_4}\sin^2\theta_o}\left(2 - \dfrac{3}{\sqrt{\sin^2\alpha + I_4\cos^2\alpha}}\right. \\ \quad \left. + \dfrac{1}{(\sin^2\alpha + I_4\cos^2\alpha)^{3/2}}\right), & 1 < I_4 < \lambda^{*2} \\ \phi\dfrac{E\cos\alpha}{2\sqrt{I_4}}\left(\beta - \dfrac{1}{\sqrt{\sin^2\alpha + I_4\cos^2\alpha}}\right), & I_4 > \lambda^{*2} \end{cases} \tag{64}$$

where now

$$I_4 = \zeta^{-1}\sin^2\psi + \zeta^2\cos^2\psi. \tag{65}$$

The equilibrium equations (49), in this case reduce to

$$\frac{\partial Q}{\partial r} = -\frac{\chi}{\zeta r}\sin^2\psi, \quad \frac{\partial Q}{\partial \theta} = 0, \quad \frac{\partial Q}{\partial z} = 0, \tag{66}$$

which allow an explicit expression for $Q$ to be obtained

$$Q = -\frac{\chi}{\zeta}\sin^2\psi \log r + C, \tag{67}$$

where $C$ is a constant. The value of $C$ is determined by again applying a traction-free boundary condition on $r = a$ ($T_{rr}|_{r=a} = T_{r\theta}|_{r=a} = T_{rz}|_{r=a} = 0$). The first of these leads to

$$C = (1 - \phi)\mu\zeta^{-1} + \frac{\chi}{\zeta}\sin^2\psi \log a, \tag{68}$$

and we now have explicit expressions for the diagonal components of the Cauchy stress

$$T_{rr} = \frac{\chi}{\zeta}\sin^2\psi \log\left(\frac{r}{a}\right), \quad T_{\theta\theta} = \frac{\chi}{\zeta}\sin^2\psi\left(1 + \log\left(\frac{r}{a}\right)\right), \tag{69}$$

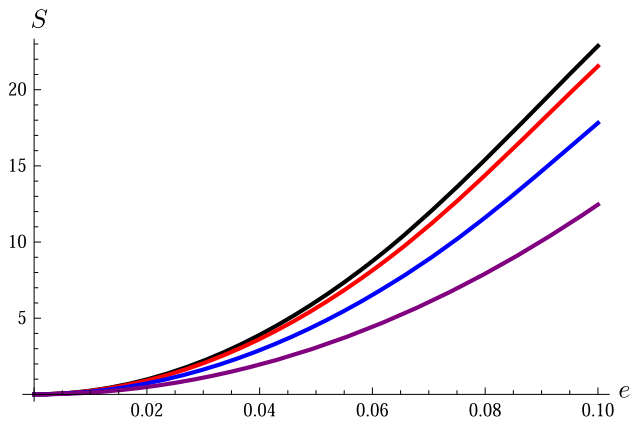

**Fig. 6.** The force per unit undeformed area as a function of the engineering strain $e = \zeta - 1$ with $\psi = \theta_0 = 20°$, $(1-\phi)\mu = 0.01$ MPa and $\phi E = 1027$ MPa. Black line: $\alpha = 0°$, red line: $\alpha = 10°$, blue line: $\alpha = 20°$, purple line: $\alpha = 30°$. (For interpretation of the references to color in this figure caption, the reader is referred to the web version of this paper.)

$$T_{zz} = (1-\phi)\mu(\zeta^2 - \zeta^{-1}) + \chi\left(\zeta^2 \cos^2\psi + \lambda^{-1}\sin^2\psi \log\left(\frac{r}{a}\right)\right). \quad (70)$$

By integrating $T_{zz}$, the resultant axial load $N$ which acts on a cross-sectional slice of the ligament or tendon can be obtained

$$N = 2\pi \int_0^a T_{zz} r \, dr$$
$$= \pi a^2 \left((1-\phi)\mu(\zeta^2 - \zeta^{-1}) + \chi\left(\zeta^2 \cos^2\psi - \frac{\sin^2\psi}{2\zeta}\right)\right). \quad (71)$$

We can use this expression to derive the average force per unit undeformed area

$$S = \frac{N}{\pi A^2} = \frac{1}{\zeta}\left((1-\phi)\mu(\zeta^2 - \zeta^{-1}) + \chi\left(\zeta^2 \cos^2\psi - \frac{\sin^2\psi}{2\zeta}\right)\right). \quad (72)$$

By varying the parameters $\alpha$, $\psi$ and $\theta_0$ in the above expression, we can determine their relative effects. In Figs. 6–8 $S$ (with $(1-\phi)\mu = 0.01$ MPa and $\phi E = 1027$ MPa, as before) is plotted as a function of the engineering strain $e = \zeta - 1$ for several values of these parameters. In each plot, we use $\alpha = \psi = \theta_0 = 20°$ as a base case (the blue lines) and investigate the effect of changing each parameter in turn by plotting the curve when that parameter is equal to $0°$ (black lines), $10°$ (red lines), and $30°$ (purple lines). In Fig. 6, $\alpha$ is varied, in Fig. 7, $\psi$ is varied, and in Fig. 8, $\theta_0$ is varied.

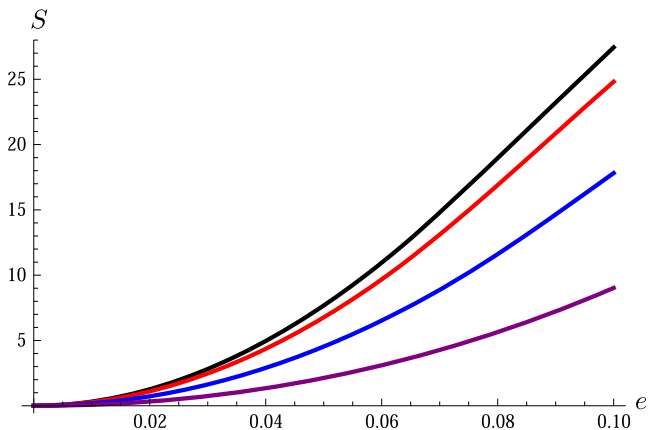

**Fig. 7.** The force per unit undeformed area as a function of the engineering strain $e = \zeta - 1$ with $\alpha = \theta_0 = 20°$, $(1-\phi)\mu = 0.01$ MPa and $\phi E = 1027$ MPa. Black line: $\psi = 0°$, red line: $\psi = 10°$, blue line: $\psi = 20°$, purple line: $\psi = 30°$. (For interpretation of the references to color in this figure caption, the reader is referred to the web version of this paper.)

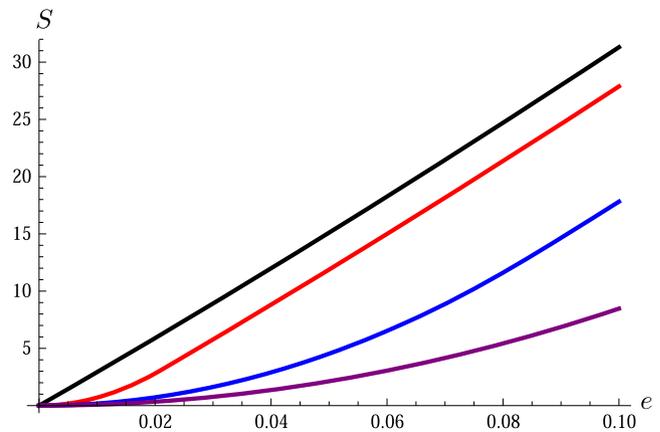

**Fig. 8.** The force per unit undeformed area as a function of the engineering strain $e = \zeta - 1$ with $\alpha = \psi = 20°$, $(1-\phi)\mu = 0.01$ MPa and $\phi E = 1027$ MPa. Black line: $\theta_0 = 0°$, red line: $\theta_0 = 10°$, blue line: $\theta_0 = 20°$, purple line: $\theta_0 = 30°$. (For interpretation of the references to color in this figure caption, the reader is referred to the web version of this paper.)

**Table 1**
The effect of varying the three parameters $\alpha$, $\psi$, $\theta_0$ on the stress at 5% strain.

|  | Parameter values | Stress (MPa) | Change (%) |
| --- | --- | --- | --- |
| Base case | $\alpha = \psi = \theta_0 = 20°$ | 4.53 | 0 |
| $\alpha$ | $0°$ | 6.09 | +34 |
|  | $10°$ | 5.66 | +25 |
|  | $30°$ | 3.07 | −32 |
| $\psi$ | $0°$ | 7.68 | +69 |
|  | $10°$ | 6.77 | +49 |
|  | $30°$ | 2.12 | −53 |
| $\theta_0$ | $0°$ | 15.1 | +233 |
|  | $10°$ | 11.9 | +162 |
|  | $30°$ | 2.12 | −53 |

Table 1 shows the stress at 5% strain when each of the parameters is varied from the base case. It appears that $\theta_0$ has the biggest effect, followed by $\psi$, then $\alpha$; however, they all clearly have a significant effect. Note that, as mentioned in Section 2, when $\theta_0 = 0°$, there is no toe-region and the ligament or tendon behaves approximately linearly. We conclude that it is the distribution of crimp angles that primarily governs the non-linear region of a ligament's or tendon's stress–strain curve and the helix angles mainly affect their stiffness.

## 6. Discussion

In this paper, a new SEF for the hyperelastic modelling of ligaments and tendons that consist of fascicles with helically arranged fibrils has been derived. This is an extension of the work of Shearer (2015) and Grytz and Meschke (2009). Grytz and Meschke included a fibril bending stiffness in their model (i.e. they modelled the fibrils as helical rods). Whilst the inclusion of this parameter may seem sensible, in their case it led to an SEF that cannot be expressed explicitly in terms of strain invariants, and instead has to be determined algorithmically (although the inclusion of bending stiffness does not always lead to an intractable SEF, Spencer and Soldatos, 2007). By neglecting bending stiffness, we have essentially modelled the fibrils as strings, which, given the extreme aspect ratios exhibited by fibrils (Trotter and Koob, 1989), seems to be a reasonable assumption. By doing this, it has been possible to incorporate fibril crimping into the model and obtain an SEF that *is* explicitly expressible in terms of strain invariants, and whose material parameters can all be directly experimentally measured. Examples of ways to measure most of these parameters are discussed in Shearer (2015). The only parameter not mentioned in that paper is the fibril helix angle, which can be measured via SEM (Yahia and Drouin, 1989).

The SEF derived here is dependent on only two invariants, but at least three are needed to fully capture the mechanical response of incompressible, transversely isotropic materials under complex loading conditions (Destrade et al., 2013). A natural extension of the SEF derived here could be obtained by replacing the neo-Hookean matrix component with a Mooney–Rivlin form.

The errors in fitting the experimental data in Section 4.1 are actually slightly higher than those achieved by the original model proposed by Shearer (2015) ($\bar{\delta} = 5.3\%$, $\bar{\Delta} = 0.12$ MPa). This may indicate that the fibril helix angle in human patellar tendon is closer to $0°$ than the $27°$ that was assumed based on images of canine patellar tendon. To fully test the model, all of its parameters should be independently measured for a specific ligament or tendon and used to predict its stress–strain behaviour. This prediction should then be compared with tension tests performed on that ligament or tendon.

In Section 4.2, it was shown that the model can be used to make qualitative predictions about how tendons respond to simple shear without having to dissect them, and therefore potentially damage their ultrastructure. Interestingly, the stress required to apply simple shear appears to be much smaller than that required for extension. This could have potential implications for tendon reconstruction surgery. If graft tendons are fixated such that shear forces are generated, large deformations could occur, which could potentially lead to rerupture. This situation could be avoided by assessing different potential graft geometries via finite element analysis. Since the model is expressed in terms of an SEF, it would be straightforward to implement it into finite element software to simulate in vivo deformations. The microstructural basis of the model makes it preferable to the phenomenonlogical finite strain models that have been implemented in commercial finite element packages to date.

To conclude, we consider the results of Section 5. All the angle parameters ($\alpha$, $\psi$, $\theta_o$) have a significant effect on the predicted mechanical properties. This may indicate that the fibrilar and fascicular structure of a ligament or tendon is optimised to suit its function. For example, one would expect to see fewer helical substructures in a stiff, positional tendon and more in those that are required to be more flexible. This prediction is supported by the experimental observations of Thorpe et al. (2012, 2013).

## Conflict of interest statement

The author has no conflicts of interest to declare.

## Acknowledgements

The author would like to thank EPSRC for funding this work through Grant EP/L017997/1 and Dr W.J. Parnell for his feedback.